\documentclass{ws-rv9x6}
\usepackage[hang]{subfigure}   % required only when side-by-side / subfigures are used
\usepackage{ws-rv-van}   % numbered citation & references (default)
\usepackage[utf8]{inputenc}
\makeindex

\begin{document}

\chapter[DPD sum rules in QCD]{DPD sum rules in QCD}
\label{ch:DPD_sum_rules_in_QCD}

\author[P. Plößl]{Peter Plößl}

\address{Institut für Theoretische Physik, Universität Regensburg, \\
93040 Regensburg, Germany,\\
peter.ploessl@physik.uni-regensburg.de}

\begin{abstract}
We review the double parton distribution (DPD) sum rules and establish their 
validity to all orders in QCD. This is done using a diagrammatic approach and 
light-front perturbation theory. In the process we furthermore investigate the 
QCD evolution of DPDs and obtain sum rules for $1\to2$ splitting kernels in 
close analogy to the DPD sum rules themselves.
\end{abstract}

\body
\section[Preliminaries]{Preliminaries}
\label{sec:Preliminaries}
Many current phenomenological studies of double parton scattering (DPS) 
rely on very simple approximations to the factorised DPS cross section using the 
pocket formula 
$
	\sigma_{(A,B)}^D
	=			
	\sigma_{A}^{S}\sigma_{B}^{S}
	/
	\sigma_{\text{eff}}
	,\,
$
which approximates the DPS cross section as the product of two single parton 
scattering (SPS) cross section divided by the supposedly process independent 
effective cross section $\sigma_{\text{eff}}$. The assumption that gives rise to 
such a form of the DPS cross section is that DPDs can be approximated as simple 
products of the well known parton distribution functions (PDFs), neglecting all 
correlations between the partons inside the hadron. However, we know that this 
approximation must fail for large momentum fractions $x_{i}$ due to momentum 
conservation and also for small interparton distances $y$ where the perturbative 
splitting of one parton to the two observed ones generates strong correlations. 
Correlations between partons are furthermore also found in dynamical 
models\cite{Rinaldi:2016jvu,Ceccopieri:2017oqe}. Therefore a more realistic 
ansatz for DPDs is needed which is however a difficult task for which any 
constraint is helpful. One possible way to constrain DPDs is provided by the DPD 
sum rules postulated by Gaunt and Stirling\cite{Gaunt:2009re} which is what 
motivated us to prove that the DPD sum rules which were derived with the parton 
model in mind are actually valid to all orders in QCD. \\
Before giving the explicit form of the some rules a short comment on the DPDs in 
these sum rules is in place. Starting from the position space DPD 
$
	F^{j_{1}j_{2}}
	\left(
		x_{i},\boldsymbol{y};\mu
	\right)\,
$
with $x_{i}=x_{1},x_{2}$ which can be interpreted as the probability density to 
find two partons of flavour $j_{1}$ and $j_{2}$, momentum fractions $x_{1}$ and 
$x_{2}$ respectively with an interparton distance $y$. The related momentum 
space DPD is as usual obtained by Fourier transforming, i.e.\
\begin{align} 
	F^{j_{1}j_{2}}
	\left(
		x_{i},\boldsymbol{\Delta};\mu
	\right)
	=
	\int\mathrm{d}^{2}\boldsymbol{y}\,
	\text{e}^{i\boldsymbol{y}\boldsymbol{\Delta}}
	F^{j_{1}j_{2}}
	\left(
		x_{i},\boldsymbol{y};\mu
	\right)\,.
\end{align}
In fact, this relation requires additional ultraviolet renormalisation, as we 
will explain below. In the sum rules these momentum space distributions occur 
evaluated at $\boldsymbol{\Delta}=0$ which corresponds to integrating the 
position space DPD over all $\boldsymbol{y}$ such that this gives the integrated 
probability to find partons $j_{1}$ and $j_{2}$ with momentum fractions $x_{1}$ 
and $x_{2}$ respectively.\\
The sum rules Gaunt and Stirling postulated are:\\
\underline{valence quark number sum rule:}
\begin{align}
	\int\limits_0^{1-x_1}\!\!\!
  \mathrm{d}x_2\,
  F^{j_1j_{2,v}}(x_i;\mu)
  &
  =
  \left(
    N_{j_{2,v}}
    +
    \delta_{j_1,\overline{j_2}}
    -
    \delta_{j_1,j_2}
  \right) 
  f^{j_1}(x_1;\mu)\,,
  \label{eq:numsum}
\end{align}
\underline{momentum sum rule:}
\begin{align}
  \sum_{j_2}\!\!\!
  \int\limits_0^{1-x_1}\!\!\!
  \mathrm{d}x_2\,x_2\, 
  F^{{j_1}{j_2}}(x_i;\mu)
  &
  =
  (1-x_1)f^{j_1}(x_1;\mu)\,,
  \label{eq:mtmsum}
\end{align}
where the valence DPD $F^{j_1j_{2,v}}$ is given by 
$F^{j_1j_{2}}-F^{j_1\overline{j_{2}}}$. 

\section[Outline of a proof for bare distributions]{Outline of a proof for bare 
distributions}
\begin{figure}[!t]
    \centering
    \subfigure[general LCPT PDF graph\label{fig:PDF-LCPT}]
    {\includegraphics[width=0.4\textwidth]{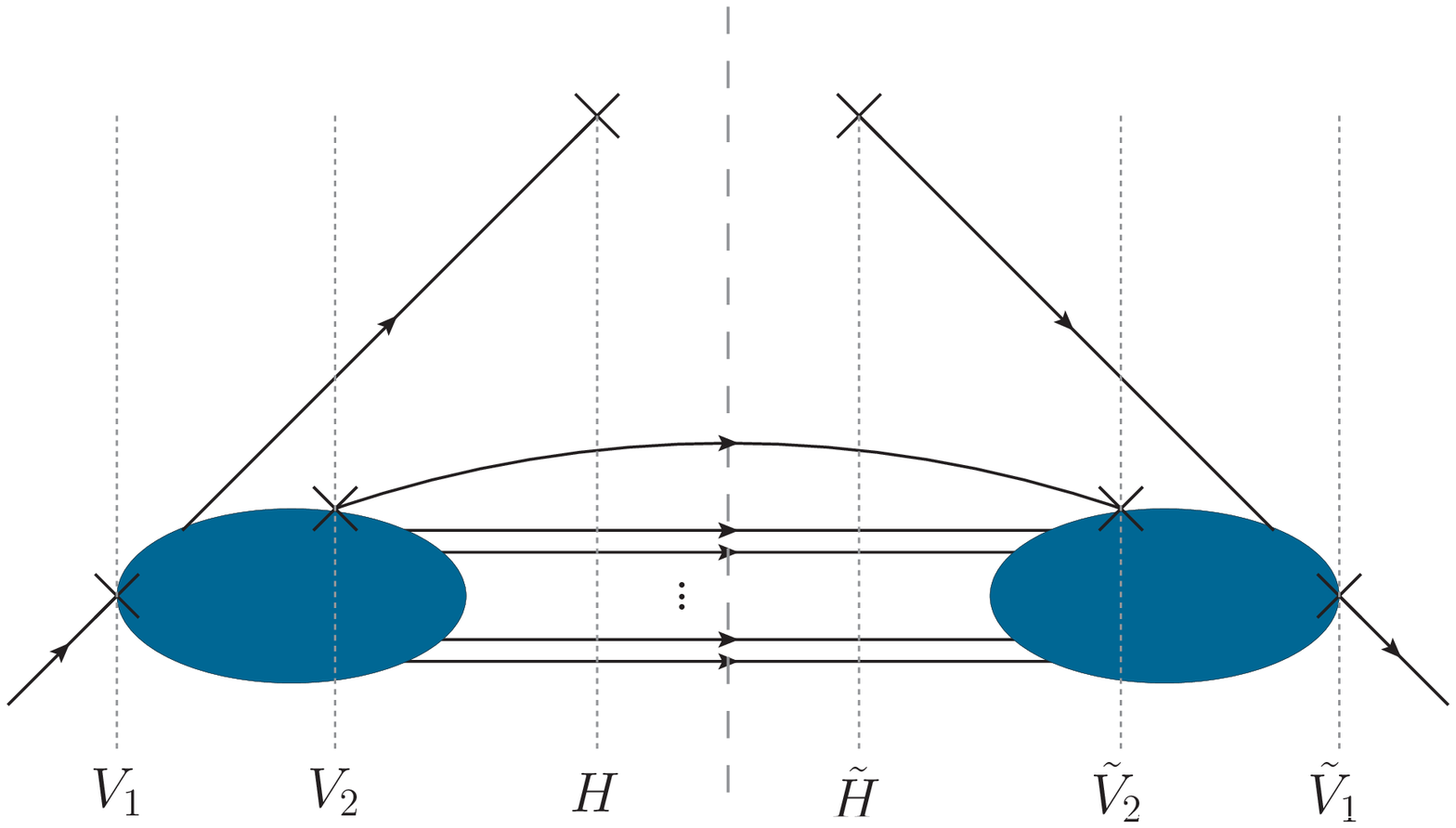}}
    \qquad
		\subfigure[corresponding LCPT DPD graph\label{fig:DPD-LCPT}]
		{\includegraphics[width=0.4\textwidth]{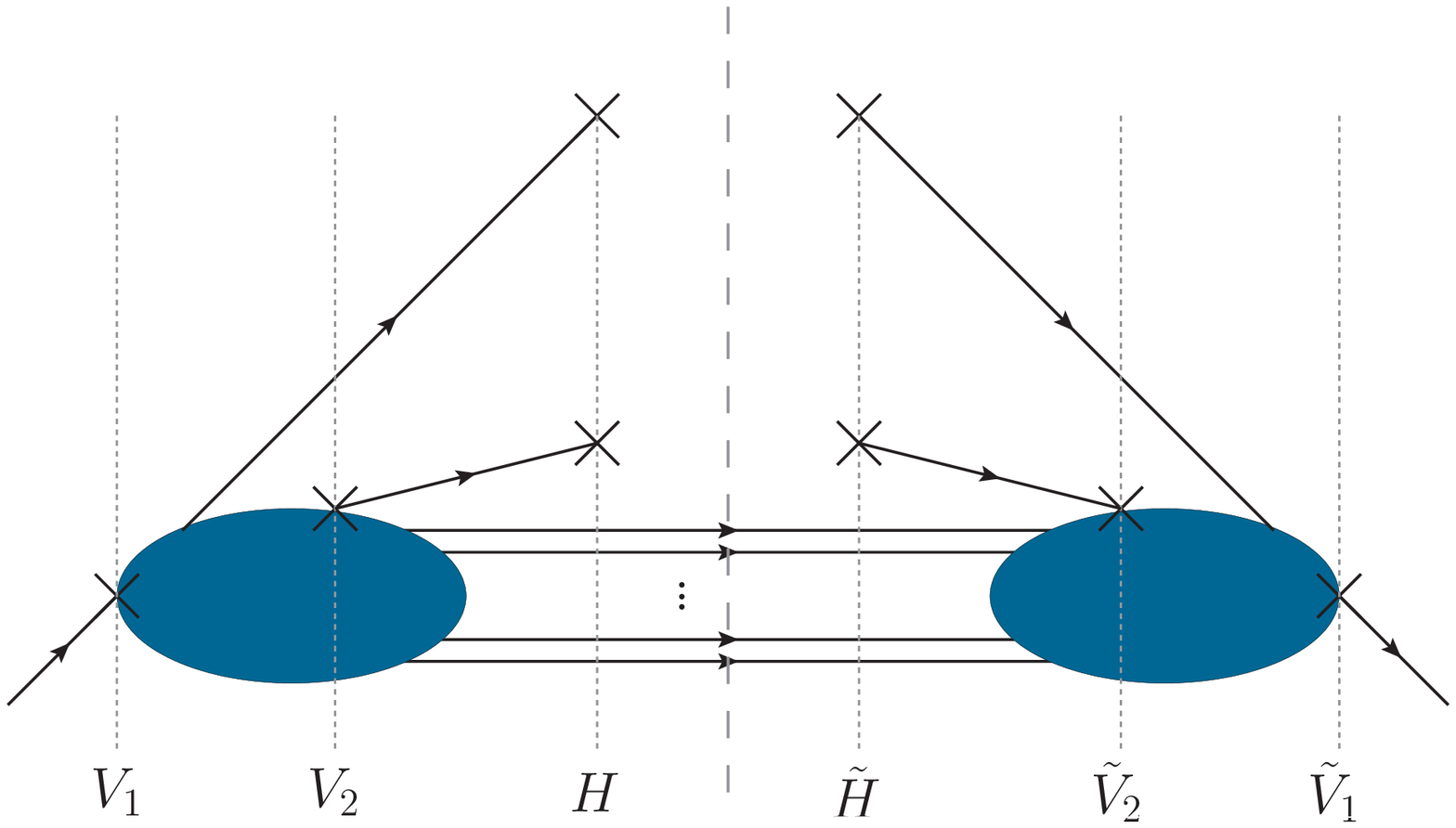}}
    \caption{transition from a given LCPT PDF graph to a corresponding LCPT DPD
    graph}
    \label{fig:PDF->DPD}
\end{figure}

We now sketch how to prove that the DPD sum rules retain their validity when 
considered in QCD with a more thorough treatment to be given in a forthcoming 
paper\cite{Diehl:2017ajj}. Earlier studies of the DPD sum rules can be found in 
appendix A of Ref.\ \cite{Blok:2013bpa} and appendix C of Ref.\ 
\cite{Gaunt:2012ths}. In order to perform the proof we first showed that they 
hold for unrenormalised distributions making use of the fact that parton 
distributions can be expressed in terms of Feynman diagrams. Of course we cannot 
actually calculate DPDs in perturbation theory, but we assume in our proof that 
the general properties of Feynman graphs hold also in the non-perturbative 
regime which is similar to the approach in factorisation proofs. Our analysis of 
1-loop examples made it clear that this proof is best performed in light-front 
ordered perturbation theory (LCPT), for details refer e.g.\ to chapter $7.2$ in 
Ref.\ \citen{Collins:2011zzd}. We could show that for PDF graphs and the 
corresponding DPD graphs obtained by ``cutting'' one of the final state lines in 
the PDF graph which is then treated as the second active parton the same 
light-cone orderings have to be considered, cf.\ Fig.\ \ref{fig:PDF->DPD}, 
allowing us to show the following equality
\begin{align}
  2
  \left(
    \!x_l \,p^+\! 
  \right)^{n_l}
  \mathcal{G}_{DPD}^{j_1j_2}
  =
  \mathcal{G}_{PDF}^{j_1}\,,
  \label{eq:equality-PDF-DPD}
\end{align}
relating PDF and DPD graphs. Here $\mathcal{G}_{PDF}^{j_1}$ and 
$\mathcal{G}_{DPD}^{j_1j_2}$ are the LCPT expressions for a given PDF graph and 
one of its corresponding DPD graphs as illustrated in Fig.\ \ref{fig:PDF->DPD} 
while $x_{l}$ is understood to be the the longitudinal momentum fraction of the 
``cut'' line in the same figure. With this the proof of the number sum rule 
reduces to showing the following
\begin{align}
  \sum_l
  \left(
    \delta_{f(l),\,j_2}-\delta_{f(l),\,\overline{j_2}}
  \right)
  &
  =
  \left(
    N_{{j_2}_v}+\delta_{j_1,\,\overline{j_2}}-\delta_{j_1,\,j_2}
  \right)
  \nonumber\\
  \left(
    N
    \left(
      \vphantom{\overline{j_{2}}}j_{2}
    \right)_{\mathcal{G},\,c}
    -
    N
    \left(
      \overline{j_{2}}
    \right)_{\mathcal{G},\,c}
  \right)
  &
  =
  \left(
    N_{{j_2}_v}+\delta_{j_1,\,\overline{j_2}}-\delta_{j_1,\,j_2}
  \right)
  \,.
  \label{eq:numsum-modified}
\end{align}
In this expression $N\left(\vphantom{\overline{j_{2}}}j_{2}\right)$ and 
$N\left(\overline{j_{2}}\right)$ are the number of $j_{2}$ and $\overline{j_{2}}$ 
quarks respectively running over the final state cut in the considered PDF graph. 
I.e.\ in order to show the validity of the number sum rule we simply have to 
count the number of $j_{2}$ and $\overline{j_{2}}$ quarks. As for gluons the 
very notion of a valence DPD is ill defined the case $j_{2}=g$ can be neglected. 
Besides the $j_{2}$ valence quarks we find an arbitrary number, $x$, of 
$j_{2}\overline{j_{2}}$ pairs inside of a hadron. As these additional quarks 
however always come in pairs it is possible to express $N\left(\vphantom{
\overline{j_{2}}}j_{2}\right)-N\left(\overline{j_{2}}\right)$ in terms of 
$j_{1}$, making the above equality evident. In order to show the validity of the 
momentum sum rule for bare distributions one has to show that the following 
equality holds
\begin{align}
  &
  \sum_l
  \int
  \mathrm{D}_2^{N(t)}
  \left[
    x_i
  \right]
  \mathrm{D}_1^{N(t)}
  \left[
    \boldsymbol{k}_i
  \right]
  \,
  x_l
  \,
  \mathcal{G}^{j_1}
  \left(
    \{x\},
    \{\boldsymbol{k}\}
  \right)
  \delta
  \left(
    1-\textstyle{\sum_{i}} x_i
  \right)
  \nonumber\\
  =
  &
  \left(
  	1-x_1
  \right)
  \int
  \mathrm{D}_2^{N(t)}
  \left[
    x_i
  \right]
  \mathrm{D}_1^{N(t)}
  \left[
    \boldsymbol{k}_i
  \right]
  \mathcal{G}^{j_1}
  \left(
    \{x\},
    \{\boldsymbol{k}\}
  \right)
  \delta
  \left(
    1-\textstyle{\sum_{i}} x_i
  \right)\,,
  \label{eq:mtmsum-proof}
\end{align}
where we used a shorthand notation for the integration measure
\begin{align}
	\int
  \mathrm{D}_a^b
  \left[
    x_i
  \right]
  =
  \prod_{i=a}^b
  \int_0^1
  \mathrm{d}x_i\,
  p^+\,,
  &&
  \int
  \mathrm{D}_a^b
  \left[
    \boldsymbol{k}_i
  \right]
  =
  \prod_{i=a}^b
  \int
  \frac{
  		\mathrm{d}^{D-2}\boldsymbol{k}_i
  	}
  	{
  		(2\pi)^{D-1}
  	}\,.
\end{align}
This can however easily be shown by performing the $x_{2}$ integration using the 
momentum conservation $\delta$ function, yielding
\begin{align}
  &
  \int
  \mathrm{D}_3^{N(t)}
  \left[
    x_i
  \right]
  \mathrm{D}_1^{N(t)}
  \left[
    \boldsymbol{k}_i
  \right]
  \left(
    1-x_1-\textstyle{\sum_{i\neq2}}x_i+\textstyle{\sum_{l\neq2}}x_j
  \right)
  \mathcal{G}^{j_1}
  \left(
    \{x\},
    \{\boldsymbol{k}\}
  \right)    
  \nonumber\\
  =
  &
  \left(
    1-x_1
  \right)
  \int
  \mathrm{D}_3^{N(t)}
  \left[
    x_i
  \right]
  \mathrm{D}_1^{N(t)}
  \left[
    \boldsymbol{k}_i
  \right]
  \mathcal{G}_{PDF_{c}}^{j_1}
  \left(
      \{x\},
      \{\boldsymbol{k}\}
  \right) \,.
  \label{eq:mtmsum-equivalence}
\end{align}
At the level of bare distributions the analysis of LCPT graphs thus fully 
confirms the parton model intuition, leaving any possible violations of the 
sum rules to be due to renormalisation effects which we considered next.

\section[Renormalisation]{Renormalisation}
The renormalised distributions are obtained from the bare ones by a convolution 
with a PDF renormalisation $Z$ factor for each twist-2 operator in the matrix 
element defining the PDFs and DPDs. For the DPD in transverse momentum space one 
finds in addition to this furthermore an inhomogeneous term needed to 
renormalise the perturbative $1\to2$ splitting, cf.\ section $3.2$ in Ref.\ 
\citen{Diehl:2017kgu}.
\begin{align}
  \sum_{i_1}\!\!
  \int\limits_{x_1+x_2}^{1}\!\!
  \frac{
  		\mathrm{d}z_{1}
  	}
  	{
  		z_1^2
  	}
  Z_{i_1,j_1j_2}
  \left(
    \frac{
    		x_{1}
    	}
    	{
    		z_1
    	},
    \frac{
    		x_{2}
    	}
    	{
    		z_1
    	};\mu
  \right)
  f_B^{i_1}
  \left(
    z_1
  \right)
  \label{eq:DPD-renormalised}\,.
\end{align}
In the minimal subtraction scheme the renormalisation factors are a series of 
pure poles in the dimensional parameter $\varepsilon$. In order to show the 
validity of the sum rules for renormalised quantities we subtract the r.h.s.\ of 
the respective sum rule from the l.h.s.\ and express the renormalised 
distributions in terms of bare ones convoluted with renormalisation factors. For 
the sum rules to hold this difference has so vanish. As both the l.h.s.\ and 
r.h.s.\ of the sum rules are finite for $\varepsilon\to0$ (they involve 
renormalised quantities, after all) we can conclude that in the difference 
between both sides all poles in $\varepsilon$ have to cancel, leaving at most a
finite difference. For both sum rules this difference can be brought to the 
following form
\begin{align}
  \sum_{i_1'}\!
  \int\limits_{x_1}^1\!
  \frac{
  		\mathrm{d}u_1
  	}
  	{
  		u_1
  	}
  f^{i_1'}
  \left(
    u_1;\mu
  \right)
  R
  \left(
    x_1,u_1;\mu
  \right)\,,
  \label{eq:lhs-rhs_numsum3}
\end{align}
where $R\left(x_1,u_1;\mu\right)$ is a function of the renormalisation factors 
$Z$ and the only possible finite contribution is due to the tree-level term of 
the PDF renormalisation factors. One finds however that the tree level terms 
vanish explicitly. This argument can also be adapted to hold in the
$\overline{\text{MS}}$ scheme. From the fact that $R$ in Eq.\ 
\eqref{eq:lhs-rhs_numsum3} vanishes we can furthermore obtain number and 
momentum sum rules for the $1\to2$ renormalisation factor $Z_{i,jk}$, namely
\begin{align}
  \int\limits_0^{1-x_1}\!\!\!
  \mathrm{d}x_2
  \left(\!
    Z_{i,jk}
    \left(
      x_{i};\mu
    \right)
    -
    Z_{i,j\overline{k}}
    \left(
      x_{i};\mu
    \right)\! 
  \right)
  &
  =
  \left(\!
    \delta_{i,k}-\delta_{i,\overline{k}}+\delta_{j,\overline{k}}
    -\delta_{j,k}\!
  \right)\!
  Z_{i,j}
  \left(
    x_1;\mu
  \right)
  \,,
  \nonumber\\
  \sum_{k}\!\!
  \int\limits_0^{1-x_1}\!\!
  \mathrm{d}x_2\,x_2\,
  Z_{i,jk}
  \left(
    x_1,x_2;\mu
  \right)
  &
  =
  \left(
    1-x_1
  \right)
  Z_{i,j}
  \left(
    x_1;\mu
  \right)\,.
  \label{eq:sum_rules_renormalisation_factor}
\end{align}

\section[Evolution]{Evolution}
The consistency of the DPD sum rules with the LO evolution was already noted in 
Ref.\ \citen{Gaunt:2009re}. As we did not make any assumptions about the 
renormalisation scale $\mu$ in the proof of the sum rules they are valid for 
all values of $\mu$ implying their stability under QCD evolution to all orders. 
We furthermore generalised the double DGLAP (dDGLAP) 
equation\cite{Shelest:1982dg,SHELEST1982325,Ceccopieri:2010kg,Ceccopieri:2014ufa} 
to higher orders and checked that the result is consistent with the stability of 
the sum rules. We find that the inhomogeneous term becomes a convolution of a 
single PDF with a $1\to2$ splitting kernel
\begin{align}
  \sum_{i_1}
  \int\limits_{x_1+x_2}^1
  \frac{
  		\mathrm{d}v
  	}
  	{
  		v^2
  	}
  P_{i_1,j_1j_2}
  \left(
    \frac{
    		x_{1}
    	}
    	{
    		v
    	},
    \frac{
    		x_{2}
    	}
    	{
    		v
    	}
  \right)
  f^{i_1}
  \left(
    v;\mu
  \right)\,,
  \label{eq:dDGLAP_higher_orders}
\end{align}
where the higher order $1\to2$ splitting kernel $P_{i,jk}$ is -- in MS -- given 
by
\begin{align}
	P_{i,jk}
	\left(
		x_{i};\alpha_{s}
		\left(
			\mu
		\right)
	\right)
	=
	-\alpha_{s}
	\left(
		\mu
	\right)
	\frac{
			\partial
		}
		{
			\partial\alpha_{s}
			% \left(
			% 	\mu
			% \right)
		}
	Z_{i,jk}^{(-1)}
	\left(
		x_{i},\alpha_{s}
		\left(
			\mu
		\right)
	\right)\,.
	\label{eq:splitting_kernel_renormalisation_factor}
\end{align}
Here $Z_{i,jk}^{(-1)}$ is the coefficient of the ${}^{1}\!/\!{}_{\varepsilon}$ 
pole of $Z_{i,jk}$. Again, this can also be adapted to $\overline{\text{MS}}$. 
A first consistency check is that the renormalisation scale dependence of 
$Z_{i,jk}$ is also governed by the dDGLAP equation as one would expect. From 
this in combination with the sum rules for the $1\to2$ renormalisation factor 
we furthermore derived analogous sum rules for the $1\to2$ splitting kernels
\begin{align}
	\int\limits_0^{1-x_1}\!\!\!
	\mathrm{d}x_2
	\left(
    P_{i,jk}
    \left(
      x_{i}
    \right)
    -
    P_{i,j\overline{k}}
    \left(
      x_{i}
    \right)
	\right)
	&
	=
	\left(
    \delta_{i,k}-\delta_{i,\overline{k}}+\delta_{j,\overline{k}}
    -\delta_{j,k}
	\right)
	P_{i,j}
	\left(
    x_1
	\right)\,,
	\nonumber\\
	\sum_{k}\!\!
  \int\limits_0^{1-x_1}\!\!
  \mathrm{d}x_2\,x_2\,
  P_{i,jk}
  \left(
    x_{i}
  \right)
  &
  =
  \left(
    1-x_1
  \right)
  P_{i,j}
  \left(
    x_1
  \right)\,.
  \label{eq:mtmsum_splitting_kernels}
\end{align}
These sum rules provide a valuable cross check for future higher order 
calculations of the $1\to2$ splitting kernels. We furthermore note that at LO 
the convolution in Eq.\ \eqref{eq:dDGLAP_higher_orders} can be performed 
trivially as the LO $1\to2$ splitting kernel 
$
	P_{i_{1},j_{1}j_{2}}
	\left(
		x_{i}
	\right)
$ 
is proportional to 
$
	\delta
	\left(
		1-x_{1}-x_{2}
	\right),
$
reproducing the LO result\cite{Shelest:1982dg,SHELEST1982325,Ceccopieri:2010kg,
Ceccopieri:2014ufa}.

\section[Perturbative splitting in DPDs]{Perturbative splitting in DPDs}
As already mentioned in section \ref{sec:Preliminaries} perturbative splitting 
gives a sizeable contribution to the DPD for small interparton distance $y$  . 
In Ref.\ \citen{Diehl:2017kgu} an expression for this perturbative splitting 
contribution is given in Eq.\ $(3.14)$. Fourier transforming this expression to 
momentum space in $D-2$ dimensions we find that the 
${}^{1}\!/\!{}_{y^{\ldots}}$ pole generates an additional 
${}^{1}\!/\!{}_{\varepsilon}$ UV pole which has to be renormalised by the 
$Z_{i_{1},j_{1}j_{2}}$ factor appearing in Eq.\ \eqref{eq:DPD-renormalised}. 
This is the actual origin of the inhomogeneous term in the renormalised DPD and 
in the dDGLAP equation. As $Z_{i_{1},j_{1}j_{2}}$ has to cancel the UV pole in 
the Fourier transformed splitting DPD their pole structure is closely related 
which makes it possible to calculate the $1\to2$ splitting kernel 
$P_{i_{1},j_{1}j_{2}}$ from the $V_{i_{1},j_{1}j_{2}}$ kernel in Eq.\ 
$(3.15)$ in Ref.\ \citen{Diehl:2017kgu} using Eq.\ 
\eqref{eq:splitting_kernel_renormalisation_factor}.

\section[DPDs at Delta=0]{DPDs at $\boldsymbol{\Delta}=0$}
An alternative way to regularise and renormalise the splitting singularity of 
the splitting DPD is to introduce a cut-off function $\Phi$ which can also be 
used to resolve the DPS SPS double counting issue\cite{Diehl:2017kgu}. 
\begin{align}
	F_{\Phi}^{j_{1}j_{2}}
	\left(
		x_{i}, \boldsymbol{\Delta}; \mu,\nu
	\right)
	&
	=
	\int
	\mathrm{d}^{2}\boldsymbol{y}\,
	\mathrm{e}^{i\boldsymbol{y}\boldsymbol{\Delta}}\,
	\Phi
	\left(
		y\nu
	\right)
	F^{j_{1}j_{2}}
	\left(
		x_i, \boldsymbol{y}; \mu
	\right)
	\label{eq:DPD_mtm_cutoff}\,.
\end{align}
As most calculations are performed in the modified minimal subtraction scheme 
a matching between the cut-off regularised DPD and the $\overline{\text{MS}}$ 
regularised version is needed. Due to the fact that 
$
	F_{\scriptscriptstyle\overline{\text{MS}}}^{j_{1}j_{2}}
	\left(
		x_{i}, \boldsymbol{\Delta}; \mu
	\right)
$ 
and
$
	F_{\Phi}^{j_{1}j_{2}}
	\left(
		x_{i}, \boldsymbol{\Delta}; \mu,\nu
	\right)
$
differ only in how the UV divergence is regularised their difference can be 
calculated in perturbation theory and has the following 
form
\begin{align}
	\sum_{i_{1}}\!
	\int\limits_{x_{1}+x_{2}}^1\!\!
	\frac{
			\mathrm{d}v			
		}
		{
			v^{2}
		}\,\,
	U_{i_{1},j_{1}j_{2}}
	\left(
		\frac{
				x_{i}
			}
			{
				v
			},
		\alpha_{s}
		\left(
			\mu
		\right),
		\log
		\left(
			\frac{
					\nu
				}
				{
					\mu
				}
		\right)
	\right)
	f^{i_{1}}
	\left(
		v;\mu
	\right)\,,
\end{align}
where the kernel $U_{i_{1},j_{1}j_{2}}$ can again be obtained from the 
$V_{i_{1},j_{1}j_{2}}$ kernel. To leading order in $\alpha_{s}$ this matching 
has already been derived in section 7 of Ref.\ \citen{Diehl:2017kgu}. It should 
be noted that the $1\to2$ splitting kernel there matches the one in this 
publication only to $\mathcal{O}(\alpha_{s})$ and for $\varepsilon=0$.

\bibliographystyle{ws-rv-van}
\bibliography{proceedings_MPI@LHC2016}

\end{document}